\pdfoutput=1
\documentclass[%
reprint,
superscriptaddress,
bibnotes,
longbibliography,
amsmath,amssymb,
pra,
floatfix,
]{revtex4-1}
\usepackage{graphicx}
\usepackage{dcolumn}
\usepackage{bm}
\usepackage{hyperref}

\usepackage{natbib}
\usepackage{mathrsfs,mathtools,color,wasysym,dsfont,here}
\usepackage{amsthm}
\usepackage{physics}
\usepackage[all]{xy}
\usepackage{tikz}
\usepackage{amsfonts}
\usepackage{array,booktabs}
\usepackage{comment}
\usetikzlibrary{calc}
\usetikzlibrary{arrows}
\usetikzlibrary{decorations.markings,decorations.pathmorphing}
\usepackage{cases}
\usepackage{siunitx}
\usepackage{tablefootnote}

\AtBeginDocument{
\heavyrulewidth=.08em
\lightrulewidth=.05em
\cmidrulewidth=.03em
\belowrulesep=.65ex
\belowbottomsep=0pt
\aboverulesep=.4ex
\abovetopsep=0pt
\cmidrulesep=\doublerulesep
\cmidrulekern=.5em
\defaultaddspace=.5em
}

\begin{document}

\title{Quantum Scalar Spin Chirality in Coplanar Kagome Antiferromagnets}

\author{Nanse Esaki}
\email{esaki-nanse0428@g.ecc.u-tokyo.ac.jp}
\affiliation{Department of Physics, Graduate School of Science, The University of Tokyo, 7-3-1 Hongo, Tokyo 113-0033, Japan}
\affiliation{Advanced Science Research Center, Japan Atomic Energy Agency, Tokai, Ibaraki 319-1195, Japan}
\author{Gyungchoon Go}
\affiliation{Department of Physics, Korea Advanced Institute of Science and Technology, Daejeon 34141, Korea}
\author{Se Kwon Kim}
\email{sekwonkim@kaist.ac.kr}
\affiliation{Department of Physics, Korea Advanced Institute of Science and Technology, Daejeon 34141, Korea}
\date{\today}
\begin{abstract}
We theoretically demonstrate that quantum fluctuations inherent to antiferromagnets can generate scalar spin chirality at zero temperature even in coplanar ordered magnets. In a kagome antiferromagnet with coplanar ground-state spin configurations, the quantum-fluctuation-induced scalar spin chirality is shown to arise at zero temperature when an effective time-reversal-like antiunitary symmetry is broken in the Hamiltonian describing fluctuations, and a magnetic point group of the classical ground state allows for its presence. The scalar spin chirality fluctuations are shown to grow further with increasing temperature by thermally excited magnons. These scalar spin chirality fluctuations can reach a magnitude comparable to the static one predicted for noncoplanar spin structures, highlighting their physical implications in coplanar spin systems.
\end{abstract}

\maketitle

{\it Introduction.}---The scalar spin chirality (SSC), defined as a triple product of spins $\chi_{ijk} = \bm{S}_i \cdot (\bm{S}_j \times \bm{S}_k)$, has been one of the most definitive indicators of the chiral nature of spin systems \cite{nagaosa2012gauge, nagaosa2012emergent}. It characterizes the fundamental unit of noncoplanar spin structures, which are related to various phenomena in ordered magnets. For example, SSC in non-coplanar spin-textured electronic systems induces equilibrium orbital magnetism by modulating electron hopping \cite{hoffmann2015topological, dos2016chirality, hanke2016role, lux2018engineering, grytsiuk2020topological, zhang2020imprinting}. In addition, SSC endows transport phenomena in ordered magnets with chiral property: It acts as an effective magnetic field for conduction electrons and magnons, driving their transverse transport, such as the electronic Hall effect \cite{taguchi2001spin, tatara2002chirality, machida2010time} and the thermal Hall effect in magnetic insulators \cite{katsura2010theory, lee2013proposal, han2017spin}, respectively. These Hall effects can be further enhanced by the skew scattering of conduction electrons \cite{ishizuka2018spin} and phonons \cite{oh2024phonon}, originating from SSC. These SSC-driven phenomena have been observed in various frustrated magnets \cite{shindou2001orbital, taguchi2001spin, tatara2002chirality, machida2010time, katsura2010theory, lee2015thermal, nakatsuji2015large, han2017spin, kim2024thermal} and chiral magnets \cite{ishizuka2018spin, neubauer2009topological, kanazawa2011large, van2013magnetic, franz2014real, kim2019tunable, xu2024universal}, highlighting the essential roles of SSC across diverse material platforms.

Unlike the intensive past investigations on the static SSC in chiral spin states, little effort has been made to explore the fluctuation-induced SSC in spin systems with collinear and coplanar ground states (in the following, we refer to them as collinear and coplanar spin systems, respectively), which are generally thought to preclude SSC. However, a recent study has demonstrated that certain collinear spin systems with no SSC at zero temperature can exhibit SSC at finite temperatures induced by thermally excited magnons (magnon-induced SSC) when Dzyaloshinskii-Moriya (DM) interactions break an effective time-reversal symmetry (TRS) \cite{Esaki2025Magnon}. In addition, quantum fluctuations are known to induce SSC in spin-disordered systems such as chiral spin liquids \cite{wen1989chiral}. While these prior works address either thermal effects in ordered magnets \cite{Esaki2025Magnon} or quantum effects in disordered phases \cite{wen1989chiral}, it remains an open and fundamental question whether the zero-point quantum fluctuations can generate a finite, measurable SSC at zero temperature in a long-range ordered magnet with no classical SSC.

\begin{figure}[t]
    \centering
    \includegraphics[width=1.0\linewidth]{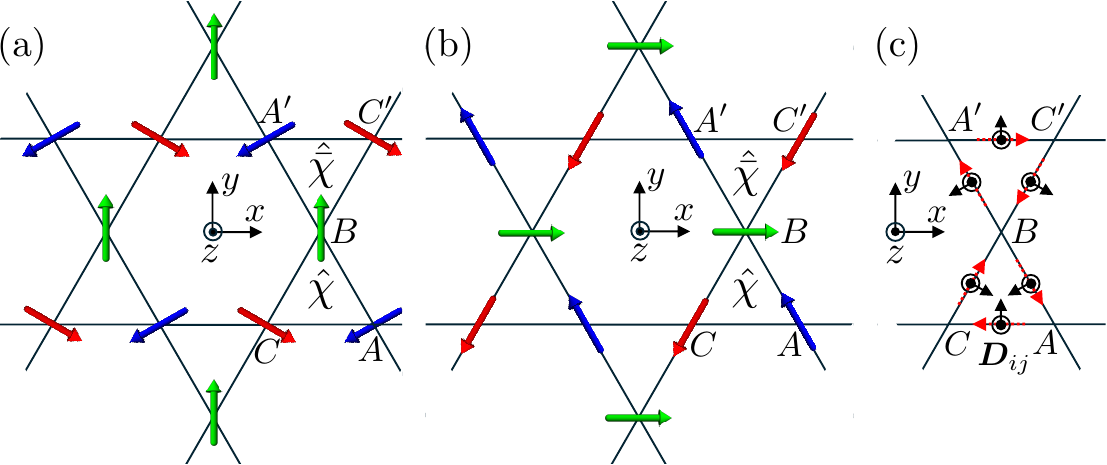}
    \caption{Coplanar ground-state spin configurations with negative vector spin chirality on the kagome lattice, parametrized by (a) $\theta = \pi/2$ and (b) $\theta = 0$, where $\theta$ is the angle of the green spin measured from the $x$-axis. The three spins in each magnetic unit cell are shown in blue, green, and red arrows, respectively. The SSC operator defined on the upper (lower) triangle formed by three sites $A, B, C$ ($A^{\prime}, B, C^{\prime}$) is denoted as $\hat{\chi}$ [Eq. (\ref{Eq:Def_chi})] ($\hat{\bar{\chi}}$). In (c), the black circles and the solid black arrows ($\hat{\bm{n}}_{ij}$) denote the directions of the out-of-plane and in-plane DM vectors, respectively. The sign of the DM vectors, $\nu_{ij}$, is taken to be positive in the direction indicated by the dashed red arrows. 
    }
    \label{fig:Model}
\end{figure}
In this Letter, we theoretically demonstrate that quantum SSC can emerge in coplanar spin systems, driven by zero-point quantum fluctuations as well as thermal excitations. Based on the linear spin-wave approximation, we present the formulation of these fluctuation-induced contributions to SSC, which we refer to as quantum and magnon-induced SSCs, respectively. Applying this formalism to a kagome antiferromagnet with a coplanar ground state in which static SSC is absent [see Fig. \ref{fig:Model}], we show that both the quantum and magnon-induced SSCs arise when an effective TRS is broken in the Hamiltonian describing fluctuations on top of the classical ground state, and a magnetic point group of the ground-state spin configuration allows for their presence. Notably, we estimate that the magnitude of both quantum and magnon-induced SSCs can be comparable to that of SSC observed in noncoplanar spin structures. Our finding of sizeable quantum- and thermal-fluctuation-induced SSC in coplanar systems unveils a previously overlooked chiral nature of ordered magnets, which calls for further investigations of fluctuation-enabled chiral phenomena that are hidden in classically nonchiral systems.

{\it Model.}---Our model system is a kagome antiferromagnet with coplanar ground-state spin configurations [see Figs. \ref{fig:Model}(a) and (b)] that is described by the following spin Hamiltonian \cite{Elhajal2002Symmetry, Laurell2018Magnon, Mook2019Thermal, Neumann2020Orbital}
\begin{equation}\label{Eq:Model_Ham}
    \mathcal{H} = \frac{J}{2} \sum_{\ev{i,j}} \bm{S}_i \cdot \bm{S}_j + \frac{1}{2} \sum_{\ev{i,j}} \bm{D}_{ij} \cdot (\bm{S}_i \times \bm{S}_j),
\end{equation}
where $\ev{i,j}$ denotes the summation over the nearest neighbor sites $i$ and $j$, $J > 0$ describes the antiferromagnetic Heisenberg interaction, and $\bm{D}_{ij}$ represents the DM vector. The DM vector in this model can be expressed as $\bm{D}_{ij} = \nu_{ij} (D_{||}\hat{n}^x_{ij}, D_{||}\hat{n}^y_{ij}, D_z)$, where $D_z$ and $D_{||}$ are the strengths of the out-of-plane and in-plane DM vectors, respectively \cite{Elhajal2002Symmetry, Laurell2018Magnon, Mook2019Thermal, Neumann2020Orbital}. The unit vector $\hat{\bm{n}}_{ij}$ denotes the direction of the in-plane DM vector, which is orthogonal to the bonds, and $\nu_{ij} = \pm 1$ describes the sign convention of the DM vectors. In Fig. \ref{fig:Model}(c), both $\hat{\bm{n}}_{ij}$ and $\nu_{ij}$ are illustrated. We note that breaking a mirror symmetry about the kagome plane is necessary for $D_{||}\neq 0$ \cite{Essafi2017Generic}, which is usually expected in real materials.

In the present study, we focus on coplanar ground-state spin configurations with negative vector spin chirality \cite{Mook2019Thermal, Neumann2020Orbital}, as shown in Figs. \ref{fig:Model}(a) and (b). Such classical ground states are stable for $D_z < 0$ and $\abs{D_{||}}$ below a critical value \cite{Elhajal2002Symmetry, Laurell2018Magnon, Mook2019Thermal, Neumann2020Orbital}. In this regime, the classical ground states with the negative vector chiral order are infinitely degenerate, all of which are related by a uniform in-plane rotation of the constituent spins. These degenerate states can be parametrized \cite{Mook2019Thermal, Neumann2020Orbital} by an angle $\theta$, denoting the angle of the green spin in Fig. \ref{fig:Model} with the $x$ direction [$\theta = \pi/2$ in Fig. \ref{fig:Model}(a) and $\theta = 0$ in Fig. \ref{fig:Model}(b)]. These accidental degeneracies of the classical states are not protected by the symmetry of the spin Hamiltonian and thus can be lifted by quantum or thermal fluctuations via the order-by-disorder mechanism, stabilizing a subset of configurations within the classically degenerate manifold \cite{villain1980order, shender1982antiferromagnetic, henley1989ordering, Rau2018PseudoGoldstone}. In the present case, the quantum correction prefers the set of sixfold degenerate ground states $\theta = \pi/2 + p\cdot \pi/3$ with $p\in \mathbb{Z}$, which is imposed by a $\mathcal{T} \times C_3(z)$ symmetry of the spin Hamiltonian (\ref{Eq:Model_Ham}) \cite{Mook2019Thermal, Neumann2020Orbital}. Here, $\mathcal{T}$ and $C_3(z)$ represent TRS and a threefold rotation about the $z$-axis, respectively.

As shown in Fig. \ref{fig:Model}, we can define two types of SSC operators on a kagome lattice, $\hat{\chi}$ and $\hat{\overline{\chi}}$, which are defined as the counterclockwise triple product of spins on the upper ($\triangle$) and lower ($\bigtriangledown$) triangles formed by the nearest-neighbor sites. Below, we shall focus only on $\hat{\chi}$ without loss of generality because $\hat{\bar{\chi}}$ is related to $\hat{\chi}$ by the spatial inversion symmetry and therefore all the results for $\hat{\chi}$ can be translated straightforwardly to those for $\hat{\bar{\chi}}$  \cite{go2024scalar, Esaki2025Magnon}. The explicit expression of $\hat{\chi}$ is given by
\begin{equation}\label{Eq:Def_chi}
    \hat{\chi} = \frac{1}{N_{\mathrm{cell}}} \sum_{A, B, C \in \triangle} \frac{\bm{S}_A \cdot (\bm{S}_B \times \bm{S}_C)}{S^3},
\end{equation}
where $N_{\mathrm{cell}}$ is the number of unit cells, $S$ the spin quantum number, and the three spin sites $A$, $B$, and $C$ belong to the three sublattices $1$, $2$, and $3$, respectively, in the same unit cell.

{\it General formalism.}---We use the linear spin-wave approximation to study SSC fluctuations by assuming sufficiently large spin $S$ (see the End Matter for details). Performing the Holstein-Primakoff transformation \cite{Holstein1940Field, kubo1952spin} on top of coplanar ground states parametrized by $\theta$, the spin Hamiltonian (\ref{Eq:Model_Ham}) becomes, up to quadratic order in the magnon operators,
\begin{equation}
\label{Eq:BdG_Ham_main}
\mathcal{H} = E_{\rm cl} - \frac{1}{4}\sum_{\bm{k}} \mathrm{Tr} H_{\bm{k}} + \frac{1}{2}\sum_{\bm{k}} \bm{\Psi}^{\dag}_{\bm{k}} H_{\bm{k}} \bm{\Psi}_{\bm{k}},
\end{equation}
where $E_{\rm cl}$ describes the classical ground-state energy, $\bm{k}$ denotes the crystal momentum, and $\bm{\Psi}^{\dag}_{\bm{k}} = (a^{\dag}_{1,\bm{k}}, \cdots, a^{\dag}_{3,\bm{k}}, a_{1,-\bm{k}}, \cdots, a_{3,-\bm{k}})$ is the bosonic vector operator with $a^{\dag}_{n,\bm{k}}$ ($a_{n,\bm{k}}$) being a creation (annihilation) operator of magnons on the sublattice $n$ ($n = 1,2,3$). Using the paraunitary matrix $T_{\bm{k}}$, the Bogoliubov transformation is performed \cite{colpa1978diagonalization} to diagonalize the Hamiltonian (\ref{Eq:BdG_Ham_main})
\begin{equation}\label{Eq:Eigenproblem_BdG_T_main}
T^{\dag}_{\bm{k}} H_{\bm{k}}T_{\bm{k}}=\mathrm{diag}(\epsilon_{1,\bm{k}}, \cdots, \epsilon_{3,\bm{k}}, \epsilon_{1,-\bm{k}}, \cdots, \epsilon_{3,-\bm{k}}), 
\end{equation}
with the order of the energies $0\leq \epsilon_{1,\bm{k}} \leq \cdots \leq \epsilon_{3,\bm{k}}$. In the same way, the SSC operator (\ref{Eq:Def_chi}) is expressed in terms of the magnon operators as
\begin{equation}\label{Eq:Quadratic_SSC_matrix_main}
    \hat{\chi} = \chi_{\rm cl} + \frac{1}{N_{\rm cell}}\sum_{\bm{k}} \Bigl[-\frac{1}{4} \mathrm{Tr} \hat{\chi}_{\bm{k}} + \frac{1}{2}\bm{\Psi}^{\dag}_{\bm{k}} \hat{\chi}_{\bm{k}} \bm{\Psi}_{\bm{k}} \Bigr],
\end{equation}
where $\chi_{\rm cl}$ corresponds to the SSC of classical ground-state spin configurations, which vanishes in collinear or coplanar ground states. The explicit expressions of $H_{\bm{k}}$ and $\hat{\chi}_{\bm{k}}$ are given in the Supplemental Material \footnotemark[1]. Performing the Bogoliubov transformation (\ref{Eq:Eigenproblem_BdG_T_main}), we obtain the thermal expectation value of $\hat{\chi}$, which we denote by $\chi$ without hat, as
\begin{subequations}\label{Eqs:SSC_eq_formulate}
\begin{align}\label{Eq:SSC_eq_formulate}
    \chi &= \chi_{\rm cl} + \chi_{\rm q} + \chi_{\rm mag}, \\ \label{Eq:Quantum_SSC}
    \chi_{\rm q} &= \frac{1}{N_{\rm cell}} \sum_{\bm{k}} \Bigl[- \frac{1}{4}\mathrm{Tr} \hat{\chi}_{\bm{k}} + \frac{1}{2} \sum_{n=1}^{3} \chi_{n,\bm{k}}\Bigr],\\ \label{Eq:Magnon_SSC}
    \chi_{\rm mag} &= \frac{1}{N_{\rm cell}}\sum_{\bm{k}}\sum_{n=1}^{3} \rho(\epsilon_{n,\bm{k}},T) \chi_{n,\bm{k}},
\end{align}
\end{subequations}
where $\rho(\epsilon_{n,\bm{k}}, T) = 1/(e^{\beta\epsilon_{n,\bm{k}}}-1)$ is the Bose distribution function with the inverse temperature $\beta = 1/(k_B T)$. Here, we define the momentum space SSC profile $\chi_{n,\bm{k}}$ by
\begin{equation}\label{Eq:SSC_profile}
    \chi_{n,\bm{k}} = (T^{\dag}_{\bm{k}}\hat{\chi}_{\bm{k}}T_{\bm{k}})_{n,n},
\end{equation}
where $()_{n,n}$ describes the $(n,n)$ element of the matrix. The derivations of Eqs. (\ref{Eq:SSC_eq_formulate})-(\ref{Eq:Magnon_SSC}) can be found in the End Matter. Equation (\ref{Eq:SSC_eq_formulate}) shows that a finite SSC can arise even in systems with a coplanar ground-state spin configuration due to the quantum [$\chi_{\rm q}$ described by Eq. (\ref{Eq:Quantum_SSC})] and thermal [$\chi_{\rm mag}$ described by Eq. (\ref{Eq:Magnon_SSC})] fluctuations. We therefore refer to $\chi_{\rm q}$ and $\chi_{\rm mag}$ as quantum and magnon-induced SSCs, respectively. Equations (\ref{Eq:SSC_eq_formulate})-(\ref{Eq:Magnon_SSC}) constitute a central result of this work, providing a unified formulation for both quantum and magnon-induced SSCs. In the following, we exemplify this formalism using the kagome antiferromagnet model described by Eq. (\ref{Eq:Model_Ham}).

{\it Symmetry analysis.}---We show that both an effective TRS and a magnetic point group of the ground-state spin configuration play essential roles for the presence of SSC fluctuations [Eqs. (\ref{Eq:Quantum_SSC}) and (\ref{Eq:Magnon_SSC})]. Specifically, it is shown below that quantum and magnon-induced SSCs are expected to be finite for $\theta = \pi/2$ ground state with $D_{||}$ and to be zero in the absence of $D_{||}$ or for $\theta = 0$ ground state.

We first show that a broken effective TRS is necessary for inducing finite SSC fluctuations. For a coplanar spin texture in the plane normal to a direction $\hat{\bm{e}}$ [$\hat{\bm{e}} = \hat{\bm{z}}$ for the model (\ref{Eq:Model_Ham})], a combination of TRS ($\mathcal{T}: \bm{S}_j \rightarrow -\bm{S}_j$) and a spin rotation by $\pi$ about $\hat{\bm{e}}$ ($C_{\hat{\bm{e}}}$) maps the texture onto itself \cite{Chen2014Anomalous, Suzuki2017Cluster, Mook2019Thermal}. This symmetry is referred to as an effective TRS $\mathcal{T}C_{\hat{\bm{e}}}$, which is broken with $D_{||}$ in the spin Hamiltonian (\ref{Eq:Model_Ham}). In the presence of the $\mathcal{T}C_{\hat{\bm{e}}}$ symmetry, the BdG (\ref{Eq:BdG_Ham_main}) and SSC (\ref{Eq:Quadratic_SSC_matrix_main}) matrices satisfy $H_{\bm{k}} = H^{\ast}_{-\bm{k}}$ and $\hat{\chi}_{\bm{k}} = -\hat{\chi}^{\ast}_{-\bm{k}}$ \footnote{See the Supplemental Material for details.}, respectively. The SSC profile (\ref{Eq:SSC_profile}) therefore satisfies 
\begin{equation}\label{Eq:SSC_profile_effective_TRS}
    \chi_{n,\bm{k}} = - \chi^{\ast}_{n,-\bm{k}} = - \chi_{n,-\bm{k}},
\end{equation}
indicating that it is an odd function with respect to $\bm{k}$. Using Eq. (\ref{Eq:SSC_profile_effective_TRS}) and the relation $\epsilon_{n,\bm{k}} = \epsilon_{n,-\bm{k}}$ \footnotemark[1], we find that the quantum (\ref{Eq:Quantum_SSC}) and the magnon-induced (\ref{Eq:Magnon_SSC}) SSCs should vanish in the presence of the $\mathcal{T}C_{\hat{\bm{e}}}$ symmetry. We note that the coplanar ground-state spin configuration with positive vector spin chirality realized for $D_z > 0$ and $D_{||} = 0$ \cite{Elhajal2002Symmetry, Laurell2018Magnon} does not exhibit SSC fluctuations due to the presence of the $\mathcal{T}C_{\hat{\bm{e}}}$ symmetry. 

\begin{figure}[t]
    \centering
    \includegraphics[width=0.95\linewidth]{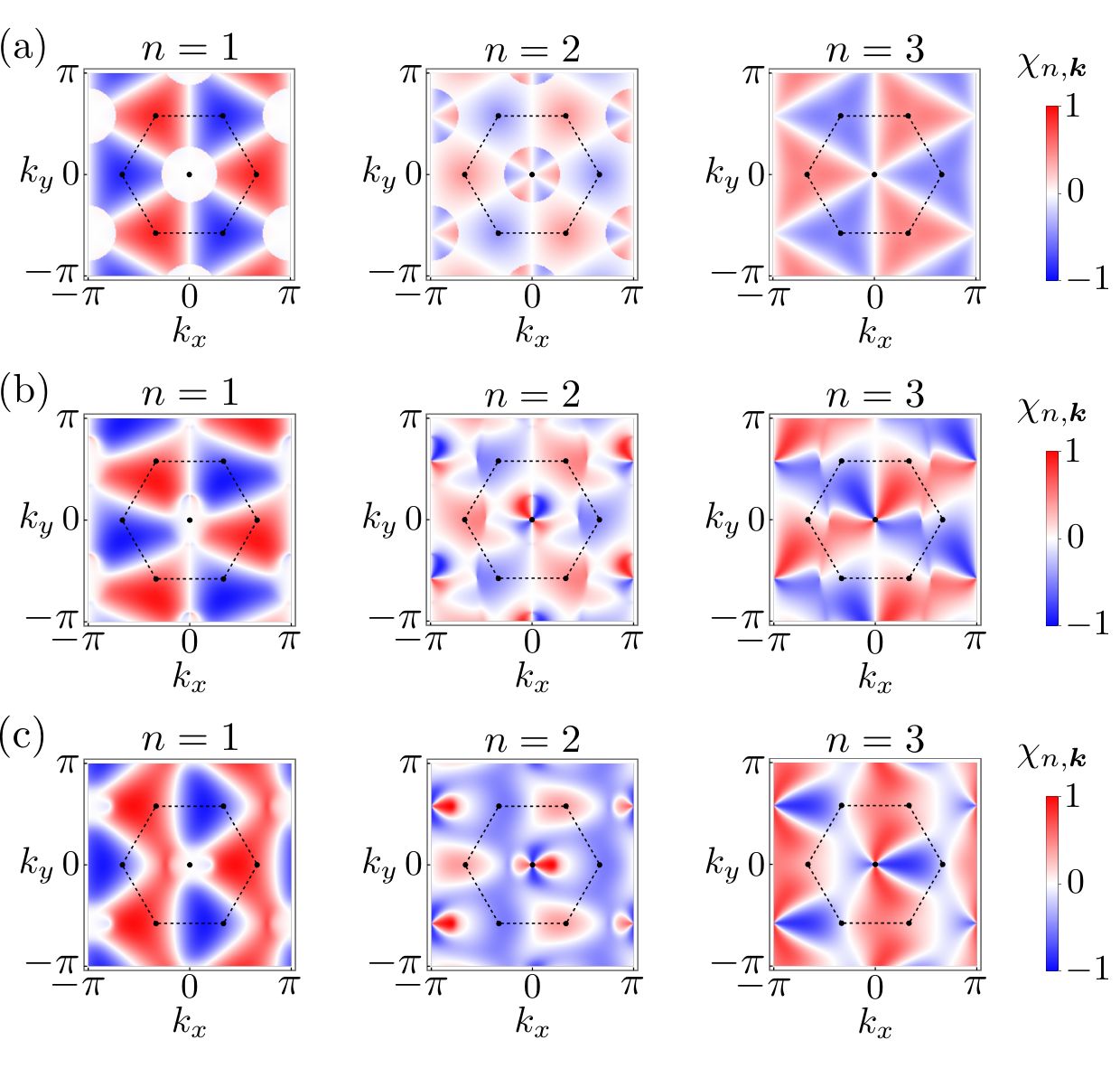}
    \caption{Plots of SSC profile $\chi_{n,\bm{k}}$ obtained by numerically calculating Eq. (\ref{Eq:SSC_profile}) for the set of parameters: $S=1, J=\SI{1}{meV}, D_{z} = -\SI{0.2}{meV}$, and (a) $D_{||} = 0$ and $\theta = \pi/2$, (b) $D_{||} = \SI{0.5}{meV}$ and $\theta = 0$, and (c) $D_{||} = \SI{0.5}{meV}$ and $\theta = \pi/2$. In each figure, the Brillouin zone is indicated by the dotted hexagon.}
    \label{fig:SSC_profile}
\end{figure}

We next show that finite SSC fluctuations can arise when allowed by the magnetic point group of the ground-state spin configuration. To demonstrate this, we analyze the symmetries of the two representative spin configurations shown in Figs. \ref{fig:Model}(a) and (b). The magnetic point group of the configuration $\theta = \pi/2$ in Fig. \ref{fig:Model}(a) is $2^{\prime}/m^{\prime}$, consisting of four symmetry elements: $E, \mathcal{T}C_2 (x), \mathcal{T}\sigma_{yz}, I$, and that of the configuration $\theta = 0$ in Fig. \ref{fig:Model}(b) is $2/m$, consisting of four symmetry elements: $E, C_2 (x), \sigma_{yz}, I$ \footnotemark[1]. Here, $C_2 (x)$ and $\sigma_{yz}$ represent a twofold rotation about an axis parallel to $\hat{\bm{x}}$ and a reflection through a mirror plane perpendicular to $\hat{\bm{x}}$, respectively. We address how these symmetries impose symmetry constraints on the BdG (\ref{Eq:BdG_Ham_main}) and SSC (\ref{Eq:Quadratic_SSC_matrix_main}) matrices in momentum space. To determine the symmetry properties of $\chi_{\bm{k}}$, it is sufficient to focus on the symmetry elements $\mathcal{T}\sigma_{yz}$ and $\sigma_{yz}$ because the other elements (except for $E$) determine the relative symmetries between $\hat{\chi}$ and $\hat{\bar{\chi}}$. In the presence of $\mathcal{T}\sigma_{yz}$, $H_{\bm{k}}$ and $\hat{\chi}_{\bm{k}}$ satisfy \footnotemark[1]
\begin{subequations}
\begin{align}\label{Eq:Hk_transform_a}
     H_{(k_x, k_y)} &=  U^{\dag}_{13} H^{\ast}_{(k_x, -k_y)} U_{13},\\ \label{Eq:chi_k_transform_a}
     \hat{\chi}_{(k_x, k_y)} & = U^{\dag}_{13} \hat{\chi}^{\ast}_{(k_x, -k_y)} U_{13},\\ \label{Eq:U_13}
     U_{13} &= \begin{pmatrix}
        1 & 0 \\
        0 & 1 \\
    \end{pmatrix} \otimes \begin{pmatrix}
        0 & 0 & 1\\
        0 & 1 & 0\\
        1& 0 & 0 \\
    \end{pmatrix}.
\end{align}
\end{subequations}
The above relations lead to
\begin{equation}\label{Eq:chi_profile_transform_a}
    \chi_{n,(k_x, k_y)} = \chi_{n, (k_x,-k_y)},
\end{equation}
and $\epsilon_{n,(k_x,k_y)} = \epsilon_{n,(k_x,-k_y)}$. Equation (\ref{Eq:chi_profile_transform_a}) shows that $\chi_{n,\bm{k}}$ is an even function with respect to $k_y$ under the $\mathcal{T}\sigma_{yz}$ symmetry. Together with Eq. (\ref{Eq:chi_profile_transform_a}) and the relation $\epsilon_{n,(k_x,k_y)} = \epsilon_{n,(k_x,-k_y)}$, we see that both the quantum (\ref{Eq:Quantum_SSC}) and magnon-induced (\ref{Eq:Magnon_SSC}) SSCs are allowed for $\theta = \pi/2$. In the same way, $H_{\bm{k}}$ and $\chi_{\bm{k}}$ satisfy \footnotemark[1]
\begin{subequations}
\begin{align}\label{Eq:Hk_transform_b}
    H_{(k_x, k_y)} &= U^{\dag}_{13} H_{(-k_x, k_y)} U_{13},\\ \label{Eq:chi_k_transform_b}
    \hat{\chi}_{(k_x, k_y)} &= -U^{\dag}_{13} \hat{\chi}_{(-k_x, k_y)} U_{13},
\end{align}
\end{subequations}
in the presence of $\sigma_{yz}$. These relations result in
\begin{equation}\label{Eq:chi_profile_transform_b}
    \chi_{n,(k_x, k_y)} = -\chi_{n, (-k_x,k_y)},
\end{equation}
and $\epsilon_{n,(k_x, k_y)} = \epsilon_{n,(-k_x, k_y)}$. Equation (\ref{Eq:chi_profile_transform_b}) indicates that $\chi_{n,\bm{k}}$ is an odd function with respect to $k_x$ under the $\sigma_{yz}$ symmetry. As a consequence, both the quantum (\ref{Eq:Quantum_SSC}) and magnon-induced (\ref{Eq:Magnon_SSC}) SSCs should vanish for $\theta = 0$.

{\it Results.}---Here, we present the numerical results of the SSC profile $\chi_{n,\bm{k}}$ in Eq. (\ref{Eq:SSC_profile}) and the SSC fluctuation $\chi$ in Eq. (\ref{Eq:SSC_eq_formulate}). Although fluctuations favor the states with $\theta = \pi/2 + p\cdot \pi/3$ ($p\in \mathbb{Z}$), any configuration within the degenerate manifold can be experimentally stabilized via a weak in-plane field. Accordingly, we treat $\theta$ as a tunable parameter, while omitting the explicit field term in Eq. (\ref{Eq:Model_Ham}) as its role is limited to state selection.

Figure \ref{fig:SSC_profile} shows the SSC profile $\chi_{n,\bm{k}}$ for various spin configurations and DM interactions. In the absence of an in-plane DM vector ($D_{||}=0$), $\chi_{n,\bm{k}}$ is an odd function of $\bm{k}$, as shown in Fig. \ref{fig:SSC_profile}(a), reflecting the presence of the effective TRS, $\mathcal{T}C_{\hat{\bm{e}}}$ [see Eq. (\ref{Eq:SSC_profile_effective_TRS})]. In contrast, a finite $D_{||}$ breaks the $\mathcal{T}C_{\hat{\bm{e}}}$ symmetry, generally allowing for a non-zero SSC. However, the magnetic point group symmetries impose specific constraints [see Eqs. (\ref{Eq:chi_profile_transform_a}) and (\ref{Eq:chi_profile_transform_b})] depending on $\theta$: for $\theta = 0$, the profile remains antisymmetric with respect to $k_x$ [Fig. \ref{fig:SSC_profile}(b)], causing the net SSC to vanish. Conversely, for $\theta = \pi/2$, such cancellation is not enforced by symmetry, resulting in a finite net SSC [Fig. \ref{fig:SSC_profile}(c)].
\begin{figure}[t]
    \centering
    \includegraphics[width=1.0\linewidth]{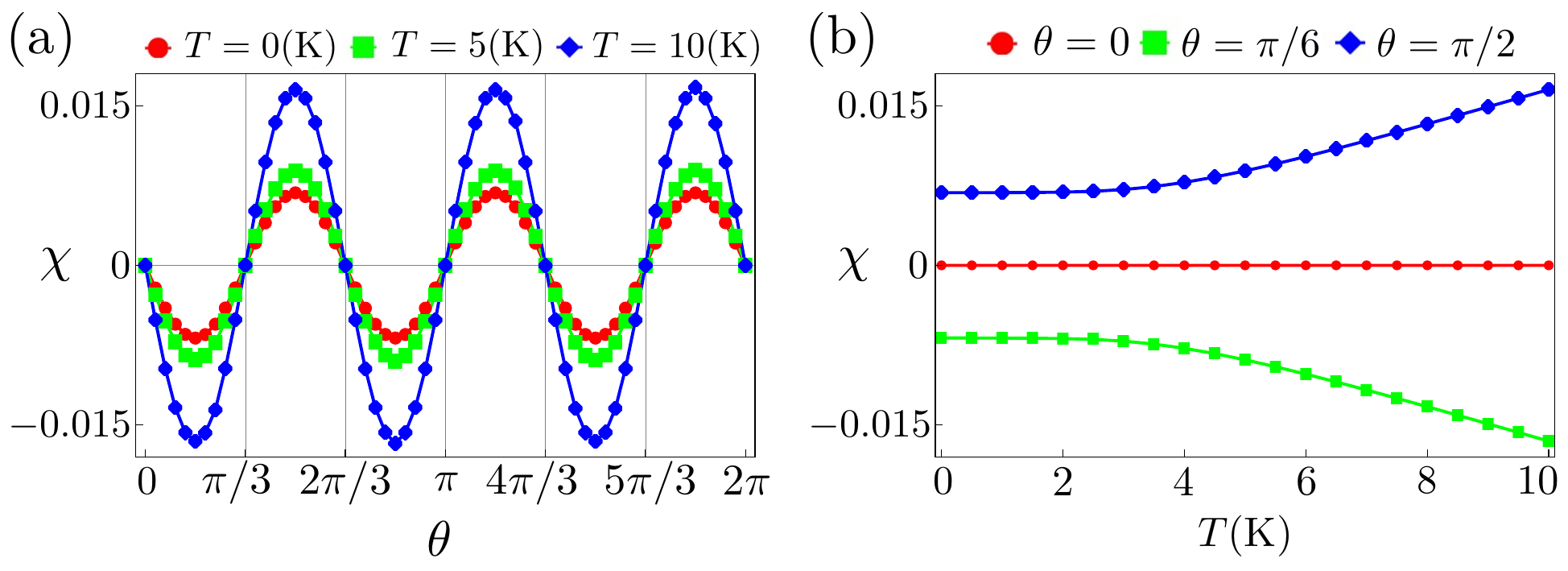}
    \caption{Plots of the expectation value of SSC, $\chi$, obtained by numerically calculating Eq. (\ref{Eq:SSC_eq_formulate}) for the set of parameters: $S=1, J=\SI{1}{meV}, D_{z} = -\SI{0.2}{meV}$, and $D_{||} = \SI{0.5}{meV}$. (a) $\theta$-dependence of $\chi$ calculated for $T = 0$, $5$, and $10$ (K). (b) Temperature dependence of $\chi$ calculated for $\theta = 0$, $\pi/6$, and $\pi/2$. The values of the SSC at $T = 0$, $\chi (T=0)$ are the quantum SSC [Eq. (\ref{Eq:Quantum_SSC})].}
    \label{fig:SSC_eq_numerical}
\end{figure}

Figure \ref{fig:SSC_eq_numerical} shows the $\theta$- and temperature- dependence of the SSC fluctuation $\chi = \chi_q + \chi_{\rm mag}$. First, the SSC fluctuation $\chi$ exhibits finite values for general coplanar states at zero temperature, i.e., the emergence of quantum SSC in coplanar kagome systems with no classical SSC, vanishing only at $\theta = 0$ aligned with the symmetry of $\chi_{n,\bm{k}}$ as shown in Fig. \ref{fig:SSC_profile}(b). Second, the SSC fluctuation $\chi$ grows further at finite temperatures, showing the contributions of thermal magnons. Third, the magnitude of the SSC fluctuation $\chi$ is maximized for the states with $\theta = \pi/2 + p \cdot \pi/3$ ($p \in \mathbb{Z}$), which are stabilized by quantum and thermal fluctuations. This figure also illustrates the following periodicities: $\chi (\theta) = \chi (\theta + p\cdot 2\pi/3)$ and $\chi (\theta) = -\chi [\theta + (2q+1)\cdot \pi/3]$ ($p,q \in \mathbb{Z}$). They originate from the $\mathcal{T} \times C_{3}(z)$ symmetry of the spin Hamiltonian (\ref{Eq:Model_Ham}) and the $C_{3}(z)$ symmetry of the SSC operator (\ref{Eq:Def_chi}), analogous to the behavior observed in the thermal Hall conductivity \cite{Mook2019Thermal}.

To assess the experimental feasibility of the SSC fluctuation, we compare our results in Fig. \ref{fig:SSC_eq_numerical} with the classical SSC in kagome antiferromagnets with noncoplanar spin configurations, $ \chi_{\rm KAFM} = \frac{3\sqrt{3}}{2} \cos^2 (\eta)\sin(\eta)$, with the canting angle $\eta$ \cite{Laurell2018Magnon}. The calculated SSC fluctuations at $T = 0$ and $T = \SI{10}{K}$ correspond to $\eta \simeq \ang{0.15}$ and $\eta \simeq \ang{0.36}$, respectively, for $\theta = \pi/2$ [see Fig. \ref{fig:SSC_eq_numerical}]. These values are within an order of magnitude of those of SSC observed in noncoplanar spin structures, which are typically on the order of $\ang{1}$ \cite{Laurell2018Magnon}, suggesting that both the quantum and magnon-induced SSCs are detectable within the current experimental reach. This highlights a key finding of our study: coplanar ground states can exhibit experimentally accessible SSC driven purely by quantum and thermal fluctuations, even though the classical ground state hosts no static SSC order.

{\it Discussion.}---In this work, we have shown theoretically that a quantum SSC emerges in coplanar kagome antiferromagnets, as well as a magnon-induced component, despite their classical ground states being free of static SSC. In our model, the in-plane component of the DM vector breaks the effective TRS and generates a $\bm{k}$-even contribution to $\chi_{n,\bm{k}}$, resulting in a finite SSC in both the quantum and thermal regimes. Such an in-plane DM component is generically expected in realistic kagome materials, where the mirror symmetry with respect to the kagome plane is broken \cite{Essafi2017Generic} by the three-dimensional crystal environment or by substrate-induced asymmetry. In addition, the angular dependence of the calculated SSCs is consistent with the symmetry arguments based on the magnetic point group of the classical ground state. The magnitude of these SSC fluctuations takes the maximum value at the angles that are selected by the order-by-disorder mechanism. We have further shown that both the quantum and magnon-induced SSCs can be comparable in magnitude to the SSC observed in noncoplanar spin structures, highlighting their experimental relevance.

We expect that our predictions of finite SSCs can be tested using experimental techniques such as neutron scattering \cite{maleyev1995investigation, maleyev1998first, simonet2012magnetic, lee2013proposal} and resonant inelastic x-ray scattering \cite{ko2011proposal, xiong2020resonant}. In particular, polarized neutron scattering can probe SSC fluctuations through the chiral component of the dynamical spin correlation function \cite{lee2013proposal, maleyev1995investigation}, which would allow one to distinguish the fluctuation-induced SSC from a small static canting of the spin structure. As an experimental platform, kagome antiferromagnets such as MgFe$_3$(OH)$_6$Cl$_2$ \cite{Fujihala2017Spin}, Cd-kapellasite CdCu$_3$(OH)$_6$(NO$_3$)$_2$ $\cdot$ H$_2$O \cite{Okuma2017Weak, Ihara2022Negative}, YCu$_3$(OH)$_6$Cl$_3$ \cite{Zorko2019Negative}, Ca-kapellasite CaCu$_3$(OH)$_6$Cl$_2$ $\cdot 0.6$ H$_2$O \cite{Ihara2020Anisotropic}, and centennialite CaCu$_3$(OD)$_6$Cl$_2$ $\cdot 0.6$ D$_2$O \cite{Iida2020q0} can be promising candidates, since they have been reported to exhibit coplanar spin structures with negative vector spin chirality.

Given the close connection of SSC to various Hall-type transport, our results indicate the potential existence of novel chiral transport phenomena in coplanar spin systems. One possibility is the Hall effect of conduction electrons \cite{taguchi2001spin, tatara2002chirality, machida2010time} induced both by the quantum and magnon-induced SSCs. In particular, our work suggests that the Hall-type phenomena can arise at zero temperature due to the quantum SSC, even with coplanar spin structures. Another possibility is the phonon thermal Hall effect originating from skew scattering \cite{oh2024phonon, Oh2025Spin} by the SSC fluctuations in coplanar spin systems. These phenomena have been primarily investigated in noncoplanar spin structures and heretofore thought to be absent in coplanar spin systems. We hope that our work triggers future research efforts toward the discovery of novel fluctuation-enabled chiral transport phenomena in otherwise nonchiral magnetic systems.

\medskip

\begin{acknowledgments}
This work was supported by Brain Pool Plus Program through the National Research Foundation of Korea funded by the Ministry of Science and ICT (2020H1D3A2A03099291) and Samsung Science and Technology Foundation (SSTF-BA2202-04). N.E. was supported by JSPS KAKENHI Grant-in-Aid for JSPS fellows Grant No. JP25KJ1167, FoPM, WINGS Program, the University of Tokyo, and JSR Fellowship, the University of Tokyo. G. G. was supported by the National Research Foundation of Korea (NRF-2022R1C1C2006578).
\end{acknowledgments}

\medskip

\section*{Data availability}
The data that support the findings of this article are available from the corresponding author upon reasonable request.

\bibliography{Ref_Quantum_SSC.bib}

\medskip

\subsection*{End Matter}
{\it Linear spin-wave theory.}---In this End Matter, we provide a general theory for the SSC fluctuations in ordered magnets based on the linear spin-wave approximation \cite{Holstein1940Field, kubo1952spin, haraldsen2009spin, petit2011numerical, toth2015linear}.

We start from a general spin interaction Hamiltonian in the quadratic form 
\begin{equation}\label{Eq:Spin_Ham_general_original}
    \mathcal{H} = \frac{1}{2}\sum_{i,j} \bm{S}^{T}_i I_{i,j} \bm{S}_j,
\end{equation}
with spin operators $\bm{S}_i$ and $\bm{S}_j$ at the lattice sites $i$ and $j$, respectively, whose magnetic interactions are comprised in the off-diagonal matrix $I_{i,j}$ ($I_{i,i} = 0$). The classical ground state of this Hamiltonian is obtained by treating the quantum mechanical spins as classical vectors and minimizing the classical ground-state energy $E_{\rm cl}$. After this procedure, the number $N_{\rm sub}$ of spins in the magnetic unit cell and the local directions of classical spin vectors $\hat{\bm{z}}_n$ are determined. We then rewrite the Hamiltonian as a sum over $N_{\rm cell}$ magnetic unit cells and a sum over $N_{\rm sub}$ magnetic sublattices as follows \cite{petit2011numerical, toth2015linear, Mook2019Thermal}
\begin{equation}\label{Eq:Spin_Ham_general}
    \mathcal{H} = \frac{1}{2} \sum_{\bm{R}_i,\bm{R}_j} \sum_{m,n} \bm{S}^T_{m,\bm{R}_i} I^{m,n}(\bm{R}_j-\bm{R}_i) \bm{S}_{n,\bm{R}_j},
\end{equation}
with each spin site being characterized by $(n,\bm{R}_j)$, where the vector $\bm{R}_j$ denotes the position of the $j$th magnetic unit cell and the index $n$ the $n$th magnetic sublattice whose position in each magnetic unit cell is given by $\bm{r}_n$. Upon the assumption of the translational invariance of the system, the interaction matrix $I^{m,n}(\bm{R}_j-\bm{R}_i)$ depends on the difference between the position vector $\bm{R}_j - \bm{R}_i$. The local coordinate system ($\hat{\bm{x}}_n, \hat{\bm{y}}_n, \hat{\bm{z}}_n$) of each spin belonging to the $n$th magnetic sublattice has to be rotated onto the global coordinate system to perform a Holstein-Primakoff transformation. Using the associated local rotation matrix
\begin{equation}\label{Eq:Rotation_matrix_general}
    R_n = (\hat{\bm{x}}_n, \hat{\bm{y}}_n, \hat{\bm{z}}_n),
\end{equation}
the transformation reads $\bm{S}_{n,\bm{R}_j} = R_n \tilde{\bm{S}}_{n,\bm{R}_j}$ for spins at the site $(n,\bm{R}_j)$. Retaining terms up to quadratic order in bosonic operators, we perform the following truncated Holstein-Primakoff transformation \cite{Holstein1940Field, kubo1952spin}:
\begin{equation}\label{Eq:HP_transform_local}
    \tilde{\bm{S}}_{n,\bm{R}_j} \simeq \begin{pmatrix}
        \sqrt{S_n/2} (a_{n,\bm{R}_j} + a^{\dag}_{n,\bm{R}_j}) \\
        -i\sqrt{S_n/2} (a_{n,\bm{R}_j} - a^{\dag}_{n,\bm{R}_j})\\
        S_n - a^{\dag}_{n,\bm{R}_j} a_{n,\bm{R}_j}
    \end{pmatrix},
\end{equation}
where $a^{\dag}_{n,\bm{R}_j}$ ($a_{n,\bm{R}_j}$) is a creation (annihilation) operator of magnons and $S_n$ is a magnitude of the classical spin vector at the site ($n, \bm{R}_j$). Within the global reference frame, this transformation becomes \cite{petit2011numerical, toth2015linear, Mook2019Thermal}
\begin{equation}\label{Eq:HP_transform_global}
    \bm{S}_{n,\bm{R}_j} \simeq \sqrt{\frac{S_n}{2}}(\hat{\bm{u}}_n a^{\dag}_{n,\bm{R}_j} + \hat{\bm{u}}^{\ast}_n a_{n,\bm{R}_j}) + \hat{\bm{z}}_n (S_n - a^{\dag}_{n,\bm{R}_j}a_{n,\bm{R}_j}),
\end{equation}
with $\hat{\bm{u}}_n = \hat{\bm{x}}_n + i\hat{\bm{y}}_n$. After Fourier transformation: $a^{\dag}_{n,\bm{R}_j} = \frac{1}{\sqrt{N_{\rm cell}}} \sum_{\bm{k}} e^{-i(\bm{r}_n+\bm{R}_j)\cdot \bm{k}} a^{\dag}_{n,\bm{k}}$, the spin Hamiltonian (\ref{Eq:Spin_Ham_general}) takes the form $\mathcal{H} = \mathcal{H}^{(0)} + \mathcal{H}^{(2)}$. Here, $\mathcal{H}^{(0)}$ is the constant term with respect to the magnon operators, and the linear term vanishes when the classical ground state is correctly obtained. The quadratic term
\begin{equation}
\label{Eq:BdG_Ham}
\mathcal{H}^{(2)} = \frac{1}{2}\sum_{\bm{k}} \bm{\Psi}^{\dag}_{\bm{k}} H_{\bm{k}} \bm{\Psi}_{\bm{k}},
\end{equation}
represents the bosonic Bogoliubov-de Gennes (BdG) Hamiltonian with the matrix $H_{\bm{k}}$ \footnotemark[1] and the bosonic vector operator 
\begin{equation}\label{Eq:vector_operator}
    \bm{\Psi}^{\dag}_{\bm{k}} = (a^{\dag}_{1,\bm{k}}, \cdots, a^{\dag}_{N_{\rm sub},\bm{k}},a_{1,-\bm{k}}, \cdots, a_{N_{\rm sub},-\bm{k}}). 
\end{equation}

The BdG Hamiltonian (\ref{Eq:BdG_Ham}) has to be diagonalized using a paraunitary matrix $T_{\bm{k}}$ \cite{colpa1978diagonalization}, which satisfies
\begin{subequations}
\begin{align}
    &T^{\dag}_{\bm{k}}\Sigma_z T_{\bm{k}} = \Sigma_z, \label{Eq:Paraunitary_matrix} \\
    &\Sigma_z = \sigma_z \otimes I_{N_{\rm sub}\times N_{\rm sub}}, \label{Eq:Sigma_z}
\end{align}
\end{subequations}
to preserve the bosonic commutation relation for the transformed operators 
\begin{equation}\label{Eq:transformed_vector_operator}
    \bm{\Psi}^{\dag}_{\bm{k}}(T^{\dag}_{\bm{k}})^{-1} = (\gamma^{\dag}_{1,\bm{k}}, \cdots, \gamma^{\dag}_{N_{\rm  sub},\bm{k}},\gamma_{1,-\bm{k}}, \cdots, \gamma_{N_{\rm sub},-\bm{k}}).
\end{equation}
Here, $\sigma_a$ is the $a = x, y, z$ component of the Pauli matrix acting on the particle-hole space, and the $N_{\rm sub}\times N_{\rm sub}$ identity matrix $I_{N_{\rm sub}\times N_{\rm sub}}$ acts on the magnetic sublattice degrees of freedom $(n = 1, \cdots, N_{\rm sub})$, respectively. Using the paraunitary matrix $T_{\bm{k}}$, the Bogoliubov transformation (\ref{Eq:Eigenproblem_BdG_T}) is performed \cite{colpa1978diagonalization}
\begin{align}\notag
T^{\dag}_{\bm{k}} H_{\bm{k}}T_{\bm{k}}&= E_{\bm{k}} \\
&=\mathrm{diag}(\epsilon_{1,\bm{k}}, \cdots, \epsilon_{N_{\rm sub},\bm{k}}, \epsilon_{1,-\bm{k}}, \cdots, \epsilon_{N_{\rm sub},-\bm{k}}), \label{Eq:Eigenproblem_BdG_T}
\end{align}
with the order of the energies $0\leq \epsilon_{1,\bm{k}} \leq \cdots \leq \epsilon_{N_{\rm sub},\bm{k}}$. This transformation diagonalizes the Hamiltonian (\ref{Eq:BdG_Ham}) using the transformed operators (\ref{Eq:transformed_vector_operator}) as
\begin{equation}\label{Eq:Diagonalization_BdG}
    \mathcal{H}^{(2)} = \sum_{\bm{k}}\sum_{n=1}^{N_{\rm sub}} \epsilon_{n,\bm{k}}\left(\gamma^{\dag}_{n,\bm{k}} \gamma_{n,\bm{k}} + \frac{1}{2}\right),
\end{equation}
which describes the low-energy excitations of the system. Acting with $T_{\bm{k}}\Sigma_z$ on the left side of Eq. (\ref{Eq:Eigenproblem_BdG_T}) and using the relation (\ref{Eq:Paraunitary_matrix}), we obtain
\begin{equation}\label{Eq:Eigenproblem_BdG}
    \Sigma_z H_{\bm{k}} T_{\bm{k}} = T_{\bm{k}} \Sigma_z E_{\bm{k}}.
\end{equation}
This relation indicates that the paraunitary matrix $T_{\bm{k}}$ can be constructed by \cite{kondo2020non}
\begin{equation}\label{Eq:T_k_explicit}
    T_{\bm{k}} = (\bm{v}_{1+,\bm{k}},\cdots,\bm{v}_{N_{\rm sub}+,\bm{k}}, \bm{v}_{1-,\bm{k}},\cdots,\bm{v}_{N_{\rm sub}-,\bm{k}}),
\end{equation}
where $\bm{v}_{n\sigma,\bm{k}}$ ($1\leq n \leq N_{\rm sub}$ and $\sigma = \pm$) is the eigenvector of $\Sigma_z H_{\bm{k}}$ corresponding to the energy $\sigma\epsilon_{n,\sigma\bm{k}}$.

{\it Formulation of scalar spin chirality fluctuations.}---The SSC operator defined on the same magnetic unit cell is given by 
\begin{equation}\label{Eq:chi_123}
    \hat{\chi} = \frac{1}{N_{\rm cell}}\sum_{\bm{R}_j} \frac{\bm{S}_{1,\bm{R}_j}\cdot (\bm{S}_{2,\bm{R}_j} \times \bm{S}_{3,\bm{R}_j})}{S_1S_2S_3}.
\end{equation}
Here and hereafter, we focus on kagome lattice systems with three sublattices, where $N_{\rm sub} = 3$. Performing the Holstein-Primakoff transformation (\ref{Eq:HP_transform_global}), we obtain
\begin{equation}\label{Eq:SSC_sum_cl_2}
    \hat{\chi} \simeq \chi_{\rm cl} + \hat{\chi}^{(2)},
\end{equation}
where we neglect the linear order in magnon operators because they do not contribute to the expectation value in thermal equilibrium. The classical SSC $\chi_{\rm cl}$ and the quadratic term $\hat{\chi}^{(2)}$ take the following form:
\begin{subequations}
\begin{align}
    \chi_{\rm cl} &= \hat{\bm{z}}_1 \cdot(\hat{\bm{z}}_2\times \hat{\bm{z}}_3) \label{Eq:Classical_SSC},\\
    \hat{\chi}^{(2)} &= \frac{1}{N_{\rm cell}} \sum_{\bm{R}_j} \sum_{l,m,n} \varepsilon_{lmn} \Bigl[-\hat{\bm{z}}_l \cdot(\hat{\bm{z}}_m\times \hat{\bm{z}}_n) \frac{1}{2S_l} a^{\dag}_{l,\bm{R}_j} a_{l,\bm{R}_j} \notag \\
    &+ \frac{1}{4\sqrt{S_m S_n}} \hat{\bm{z}}_l \cdot \bigl[(\hat{\bm{u}}_m \times \hat{\bm{u}}_n) a^{\dag}_{m,\bm{R}_j} a^{\dag}_{n,\bm{R}_j} \notag \\
    & + (\hat{\bm{u}}_m \times \hat{\bm{u}}^{\ast}_n) a^{\dag}_{m,\bm{R}_j} a_{n,\bm{R}_j} + (\rm{H.c.}) \bigr] \Bigr], \label{Eq:Quadratic_SSC}
\end{align}
\end{subequations}
where $\varepsilon_{lmn}$ is the Levi-Civita antisymmetric tensor. After Fourier transformation, the quadratic term becomes 
\begin{subequations}
\begin{align}
    \hat{\chi}^{(2)} &= \frac{1}{N_{\rm cell}}\sum_{\bm{k}} \Bigl[-\frac{1}{4} \mathrm{Tr} \hat{\chi}_{\bm{k}} + \frac{1}{2}\bm{\Psi}^{\dag}_{\bm{k}} \hat{\chi}_{\bm{k}} \bm{\Psi}_{\bm{k}} \Bigr], \label{Eq:Quadratic_SSC_matrix}\\  \label{Eq:SSC_matrix}
    \hat{\chi}_{\bm{k}} &= \begin{pmatrix} D_{\bm{k}}-F & E_{\bm{k}}\\
   E^{\ast}_{-\bm{k}} & D^{\ast}_{-\bm{k}}-F 
   \end{pmatrix},
\end{align}
\end{subequations}
The elements of the SSC matrix $\hat{\chi}_{\bm{k}}$ are given by
\begin{subequations}\label{Eq:SSC_sub_matrix}
\begin{align}\label{Eq:SSC_sub_matrix_D} 
    (D_{\bm{k}})_{m,n} &= \frac{\varepsilon_{lmn}}{2\sqrt{S_m S_n}} \hat{\bm{z}}_l \cdot (\hat{\bm{u}}_m \times \hat{\bm{u}}^{\ast}_n) e^{i(\bm{r}_n - \bm{r}_m)\cdot \bm{k}},\\
    (E_{\bm{k}})_{m,n} &= \frac{\varepsilon_{lmn}}{2\sqrt{S_m S_n}} \hat{\bm{z}}_l \cdot (\hat{\bm{u}}_m \times \hat{\bm{u}}_n) e^{i(\bm{r}_n - \bm{r}_m)\cdot \bm{k}}, \label{Eq:SSC_sub_matrix_E} \\
    (F)_{m,n} &= \frac{\delta_{m,n}}{S_m} \hat{\bm{z}}_1 \cdot(\hat{\bm{z}}_2\times \hat{\bm{z}}_3). \label{Eq:SSC_sub_matrix_F}
\end{align}
\end{subequations} 
Using the Bogoliubov transformation (\ref{Eq:transformed_vector_operator}) and the following relations \cite{Neumann2020Orbital}:
\begin{subequations}
\begin{align}
    \ev{\gamma^{\dag}_{m,\bm{k}} \gamma_{n,\bm{k}}} &= \delta_{m,n}\rho(\epsilon_{m,\bm{k}}, T)\\
    \ev{\gamma_{m,\bm{k}} \gamma^{\dag}_{n,\bm{k}}} &= \delta_{m,n}[1+\rho(\epsilon_{m,\bm{k}}, T)],\\
    \ev{\gamma^{\dag}_{m,\bm{k}} \gamma^{\dag}_{n,\bm{-k}}} &= \ev{\gamma_{m,\bm{-k}} \gamma_{n,\bm{k}}}_T = 0,
\end{align}
\end{subequations}
we derive Eqs. (\ref{Eq:SSC_eq_formulate})-(\ref{Eq:Magnon_SSC}) with Eq. (\ref{Eq:SSC_profile}). Here, $\ev{}$ denotes the thermal expectation value. We also use
\begin{equation}\label{Eq:SSC_profile_particle_hole}
    \chi_{n+N_{\rm sub},\bm{k}} = \chi^{\ast}_{n,-\bm{k}} = \chi_{n,-\bm{k}} \quad (1\leq n\leq N_{\rm sub}),
\end{equation}
which can be derived by $\bm{v}_{n-,\bm{k}} = \Sigma_x \bm{v}^{\ast}_{n+,-\bm{k}}$ and $\Sigma_x \hat{\chi}_{\bm{k}}\Sigma_x = \hat{\chi}^{\ast}_{-\bm{k}}$ with $\Sigma_x = \sigma_x \otimes I_{N_{\rm sub}\times N_{\rm sub}}$ \cite{kondo2020non} and the following relation:
\begin{equation}\label{Eq:SSC_profile_real}
    \chi_{n,\bm{k}} = (T^{T}_{\bm{k}}\hat{\chi}^{T}_{\bm{k}}T^{\ast}_{\bm{k}})_{n,n} = [(T^{\dag}_{\bm{k}}\hat{\chi}^{\dag}_{\bm{k}}T_{\bm{k}})_{n,n}]^{\ast} = \chi^{\ast}_{n,\bm{k}},
\end{equation}
which is consistent with the fact that $\chi$ (\ref{Eq:SSC_eq_formulate}) is a physical quantity, i.e., a real number.

Here, we provide a physical picture of the SSC fluctuation formulated above. Equation (\ref{Eq:Quadratic_SSC}) shows that the SSC fluctuation has a $1/S$ dependence within the present formalism. This is natural in the sense that the linear spin-wave approximation takes into account terms up to $\mathcal{O}(1/S)$ compared to the classical term (such as $E_{\rm cl}$ and $\chi_{\rm cl}$). Within this approximation, the SSC fluctuation in Eq. (\ref{Eq:Quadratic_SSC}) can be expressed in terms of vector spin chirality fluctuations as follows
\begin{align}\notag
    & \chi \simeq\frac{1}{N_{\rm cell}}\frac{1}{S_1 S_2 S_3} \sum_{\bm{R}_j} \bigl[\ev{\bm{S}_{1,\bm{R}_j}} \cdot \bigl[\ev{\bm{S}_{2,\bm{R}_j}} \times \ev{\bm{S}_{3,\bm{R}_j}} \bigr]\\ \label{Eq:SSC_fluc_rewrite}
    &+ \sum_{l,m,n} \frac{\varepsilon_{lmn} S_l \hat{\bm{z}}_l}{2} \cdot \ev{\bm{S}_{m,\bm{R}_j} \times \bm{S}_{n,\bm{R}_j}}_{\rm fluc} \bigr].
\end{align}
Here, the vector spin chirality fluctuation $\ev{\bm{S}_{m,\bm{R}_j} \times \bm{S}_{n,\bm{R}_j}}_{\rm fluc}$ is defined by
\begin{align}\notag
    &\ev{\bm{S}_{m,\bm{R}_j} \times \bm{S}_{n,\bm{R}_j}}_{\rm fluc}\\ \label{Eq:vector_chiral_fluc}
    &= \ev{\bm{S}_{m,\bm{R}_j} \times \bm{S}_{n,\bm{R}_j}} - \bigl[\ev{\bm{S}_{m,\bm{R}_j}} \times \ev{\bm{S}_{n,\bm{R}_j}}\bigr].
\end{align}
This can be understood by rewriting the expectation values in Eqs. (\ref{Eq:SSC_fluc_rewrite}) and (\ref{Eq:vector_chiral_fluc}) as
\begin{subequations}
\begin{align}\notag
    &\frac{1}{S_1 S_2 S_3}\ev{\bm{S}_{1,\bm{R}_j}} \cdot \bigl[\ev{\bm{S}_{2,\bm{R}_j}} \times \ev{\bm{S}_{3,\bm{R}_j}} \bigr] \\ \notag
    &\simeq \chi_{\rm cl}-\sum_{l,m,n} \frac{\varepsilon_{lmn}}{2S_l} \hat{\bm{z}}_l \cdot (\hat{\bm{z}}_m \times \hat{\bm{z}}_n) \ev{a^{\dag}_{l,\bm{R}_j} a_{l,\bm{R}_j}},\\ \label{Eq:Mag_triple_product_HP}
    & \frac{1}{S_m S_n} \ev{\bm{S}_{m,\bm{R}_j} \times \bm{S}_{n,\bm{R}_j}}_{\rm fluc}\\ \notag
    &\simeq \frac{1}{2\sqrt{S_m S_n}} \Bigl[(\hat{\bm{u}}_m \times \hat{\bm{u}}_n) \ev{a^{\dag}_{m,\bm{R}_j} a^{\dag}_{n,\bm{R}_j}}\\ \label{Eq:Vector_chiral_HP}
    &+ (\hat{\bm{u}}_m \times \hat{\bm{u}}^{\ast}_n) \ev{a^{\dag}_{m,\bm{R}_j} a_{n,\bm{R}_j}} + (\rm{H.c.})\Bigr],
\end{align}
\end{subequations}
respectively, where we retain terms up to $\mathcal{O}(1/S)$. Equation (\ref{Eq:SSC_fluc_rewrite}) indicates that the SSC fluctuation in coplanar spin systems with large $S$ can be understood in terms of the vector spin chirality fluctuations.

\end{document}